\documentstyle[manuscript,eqsecnum,aps,array,subeqnar]{revtex}

\begin{document}

\rightline{SNUTP01-042}

\vspace{1cm}

\centerline{\large\bf Supersymmetry-based Approach to Quantum
Particle Dynamics}

\centerline{\large\bf on a Curved Surface with Non-zero Magnetic
Field}

\vspace*{0.25cm}

\centerline { Seok Kim\footnote{Email address: calaf2@snu.ac.kr}
and Choonkyu Lee\footnote{Email address: cklee@phya.snu.ac.kr}}

\centerline{ \it Department of Physics and Center for Theoretical
Physics} \vspace{-0.25cm} \centerline{ \it Seoul National
University, Seoul 151-742, Korea}





\begin{abstract}
We present the $N=2$ supersymmetric formulation for the classical
and quantum dynamics of a nonrelativistic charged paritcle on a
curved surface in the presence of a perpendicular magnetic field.
For a particle moving on a constant-curvature surface in a
constant magnetic field, our Hamiltonian possesses the
shape-invariance property in addition. On the surface of a sphere
and also on the hyperbolic plane, we exploit the supersymmetry and
shape-invariance properties to obtain complete solutions to the
corresponding quantum mechanical problems.
\end{abstract}
\newpage




\section{Introduction}

The Landau levels and associated wavefunctions\cite{landau} are of
crucial importance in understanding many striking observations
made for two-dimensional or planar systems of electrons in a
perpendicular magnetic field, as in the cases of the de Haas-van
Alphen effect in metals\cite{kittel} and the quantum Hall
effect[3,4]. To explain the quantum Hall effect in particular,
Laughlin\cite{laugh} in his ground-breaking effort proposed
variational wave functions describing incompressible states
corresponding to special rational filling fractions of the highly
degenerate lowest Landau level. Haldane\cite{hald} then introduced
a variant of the Laughlin wave functions by considering a system
of electrons constrained to the surface of a sphere. On the
two-sphere, the fact that single-particle energy levels have
finite degeneracy only makes numerical simulations more tractable
and allows for a simplified treatment of the thermodynamic limit.
Note that the one-particle states for the latter case can also be
found in closed forms.

It is the supersymmetry that is largely responsible for the
mathematical structure of the Landau levels. This is seen by
considering the Pauli Hamiltonian for a nonrelativistic spin-$1/2$
particle in a plane,
\begin{equation}\label{flat-ham}
\begin{array}{c}
H=\frac{1}{2m}[\vec{p}-\vec{A}(\vec{x})]^{2}-\frac{1}{2m}\sigma_{3}B(\vec{x})\hspace{1.4cm}\vspace{0.2cm}\\
*\hspace{1cm}=\frac{1}{2m}\left[\sigma_{1}\left(p_{1}-A_{1}(\vec{x})\right)
+\sigma_{2}\left(p_{2}-A_{2}(\vec{x})\right)\!\frac{}{}\!\right]^{2},
\end{array}
\end{equation}
where $\vec{x}\equiv(x^{1},x^{2})$,
$B(\vec{x})=\partial_{1}A_{2}(\vec{x})-\partial_{2}A_{1}(\vec{x})$,
and $(\sigma_{1},\sigma_{2},\sigma_{3})$ are the usual Pauli
matrices. This quantum system is known to possess an $N=2$
supersymmetry[7,8], and one may exploit this supersymmetry to find
for instance the exact zero-energy ground-state wavefunctions of
the system in the presence of an arbitrary, spatially dependent,
magnetic field[9,10]. In this paper we shall generalize the $N=2$
supersymmetry property of the Pauli Hamiltonian to the case when
the underlying two-dimensional manifold is a curved surface. This
system can also be recast as an $N=2$ supersymmetric system
involving a pair of Schr\"{o}dinger Hamiltonians (i.e., for
spinless particles) defined on the curved surface.

For a complete quantum-mechanical solution, however, the existence
of supersymmetry is not enough --- we need also the
\textit{shape-invariance}[11,12]. For the Pauli system defined on
a curved surface, it turns out that shape-invariant Hamiltonians
are obtained when both the scalar curvature and the external
magnetic field (perpendicular to the surface) are restricted to be
constant. The system defined on the two-sphere or on the
hyperbolic plane(also called the pseudo-sphere) belongs to such,
and these cases provide interesting generalizations of the planar
Landau level problem. In this paper we thus find complete energy
eigenfunctions of the corresponding quantum system by exploiting
supersymmetry and shape-invariance. See Refs.[13-15] for some
previous works on these problems, but with different emphasis. The
purpose of the present work is to provide a concise, yet
self-contained, treatment of the whole problem from the
perspective of supersymmetry. A particularly comprehensive
treatment using more traditional method of mathematical physics is
given in Ref.\cite{dunne}. [Very recently, there appeared also a
preprint by G.A.Mezincescu and L.Mezincescu\cite{mezin} which
contains materials related to the present work].

This paper is organized as follows. In Section II we begin with
the Lagrangian/Hamiltonian description of the classical $N=2$
supersymmetric model of a nonrelativistic spinning particle
coupled to an external gauge field in a general two-dimensional
manifold. The corresponding quantum theory is discussed in Section
III. In the Schr\"{o}dinger picture, we here find the matrix
Schr\"{o}dinger equation which may be recognized either as a
natural curved-space generalization of the usual Pauli equation
for a spin-$1/2$ particle, or as that appropriate to a system
consisting of a pair of superpartner Hamiltonians for a spinless
particle. In Section IV we study the cases described by
shape-invariant Hamiltonians, that is, the Landau Hamiltonian on
the two-sphere or on the hyperbolic plane. For both cases, the
energy levels and complete energy eigenfunctions are produced by
our supersymmetry-based approach. Concluding remarks are made in
Section V.

\section{Classical Theory with N=2 Supersymmetry}

In this section we will discuss the classical theory of a
nonrelativistic charged spinning particle on a two-dimensional
curved surface with the metric $g_{\mu\nu}(x)$ for conveniently
chosen coordinates $x^{\mu}=(x^{1},x^{2})$. Interaction with
external vector potentials $A_{\mu}(x)\equiv(A_{1}(x),A_{2}(x))$
is included. To describe a spinning particle, it is useful to
introduce the zweibein $e^{\mu}_{a}(x)$ ($a=1,2$ refer to
components relative to the local orthonormal frame on the surface)
satisfying
\begin{equation}
e^{a}_{\mu}(x)e^{\mu}_{b}(x)=\delta^{a}_{b}\ ,\ \
e^{a}_{\mu}(x)e^{a}_{\nu}(x)=g_{\mu\nu}(x),
\end{equation}
and the corresponding spin-connection $\omega_{\mu
ab}(x)\equiv\omega_{\mu}(x)\epsilon_{ab}$ which is related to the
zweibeins in the usual manner\cite{def}.

On a curved surface, electromagnetic potentials are described by a
one-form $\textbf{A}=A_{\mu}dx^{\mu}$ and are used to define the
field strength two-form by $\textbf{F}=d\textbf{A}=
\frac{1}{2}F_{\mu\nu}dx^{\mu}\wedge dx^{\nu}$ where
$F_{\mu\nu}=\partial_{\mu}A_{\nu}-\partial_{\nu}A_{\mu}$. Using
the frame one-form $\textbf{e}^{a}=e^{a}_{\mu}(x)dx^{\mu}$, the
electromagnetic field strength is equivalently expressed as
$\textbf{F}=B(x)\textbf{e}^{1}\wedge \textbf{e}^{2}$ with
$B(x)=\frac{1}{2}\epsilon_{ab}e^{a}_{\mu}(x)e^{b}_{\nu}(x)F^{\mu\nu}(x)$,
$\epsilon_{ab}$ being a totally-skew symbol. The quantity $B(x)$
can be taken to describe the strength of the magnetic field in the
direction `normal' to the surface; this is natural since one may
represent the magnetic flux (or the first Chern number) over a
surface $M$ by $\Phi=\int_{M}\textbf{F}=\int_{M}Bd(vol)$ with the
volume form $d(vol)=\sqrt{g}\ dx^{1}\wedge
dx^{2}=\textbf{e}^{1}\wedge \textbf{e}^{2}$, where
$g=\det(g_{\mu\nu})$.

Then, taking as our dynamical variables two bosonic position
coordinates $x^{\mu}(t)$ and two real Grassmann variables (for the
spin degrees of freedom\cite{bere}) $\psi_{a}(t)$, consider the
Lagrangian\cite{fubi}
\begin{equation}\label{curved-lag}
\begin{array}{c}
L(x,\dot{x},\psi,\dot{\psi})=\frac{m}{2}g_{\mu\nu}(x)\dot{x}^{\mu}\dot{x}^{\nu}
+\frac{i}{2}\psi_{a}{\dot{\psi}}_{a}+{\dot{x}}^{\mu}{A}_{\mu}(x)\hspace{0.5cm}\vspace{0.2cm}\\
*\hspace{1cm}-\dot{x}^{\mu}\omega_{\mu}(x)S(\psi)+\frac{1}{m}B(x)S(\psi),
\end{array}
\end{equation}
where $S(\psi)\equiv-\frac{i}{2}\epsilon_{ab}{\psi}_{a}{\psi}_{b}$
represents the spin of the particle. Brief explanations on various
terms appearing in the Lagrangian (\ref{curved-lag}) might be
desirable. The first and third terms are the ones needed to
describe a spinless particle moving on a curved surface under the
action of external vector potentials. The second and fourth terms
are responsible for the spin dynamics on a curved surface ; in
fact, these two terms may be combined to yield the expression with
the covariant derivative $\frac{D}{dt}$\cite{gaume}:
\begin{equation}
\frac{i}{2}\psi_{a}({\dot{\psi}}_{a}+\dot{x}^{\mu}\omega_{\mu ab
}(x)\psi_{b})\equiv\frac{i}{2}\psi_{a}\frac{D}{dt}\psi_{a}.
\end{equation}
The last term $B(x)S(\psi)$, which represents the interaction
between the spin and external magnetic field with a very specific
gyromagnetic ratio, has been included to make the above theory be
invariant under the supersymmetry transformation
\begin{subeqnarray}
&\delta x^{\mu}=i\frac{\epsilon}{\sqrt{m}} e^{\mu}_{a}(x)\psi_{a}
\equiv i\frac{\epsilon}{\sqrt{m}}\psi^{\mu},&\slabel{x-transf-1}\\
&\delta \psi^{\mu}=-\epsilon\sqrt{m}
\dot{x}^{\mu}&\slabel{psi-transf-1}
\end{subeqnarray}
with a real Grassman parameter $\epsilon$. Note that, using
$\psi_{a}$, (\ref{psi-transf-1}) can also be written as
\begin{equation}\label{psi-transf-1'}
\delta\psi_{a}=-\frac{\epsilon}{\sqrt{m}} e_{\mu
a}(x)(m\dot{x^{\mu}}-\omega^{\mu}(x)S(\psi)).
\end{equation}

The Lagrangian invariant under the transformation (2.4a,b) exists
in general spatial dimension, but, for our Lagrangian
(\ref{curved-lag}) which applies specifically to the
two-dimensional case, we have in fact an \textit{additional}
supersymmetry. Such extended supersymmetry structure can be seen
most clearly in the Hamiltonian formulation. By the simple
Legendre transform with (\ref{curved-lag}), the Hamiltonian
appropriate to our system reads
\begin{equation}\label{curved-ham}
H=\frac{1}{2m}g^{\mu
\nu}(x)\left[p_{\mu}\!-\!A_{\mu}(x)\!+\!\omega_{\mu}(x)S(\psi)\right]
\left[p_{\nu}\!-\!A_{\nu}(x)\!+\!\omega_{\nu}(x)S(\psi)\right]-\frac{1}{m}B(x)S(\psi)
\end{equation}
with the Poisson brackets
\begin{equation}
\{x^{\mu},p_{\nu}\}_{\scriptscriptstyle{PB}}=\delta^{\mu}_{\nu}\
,\ \{\psi_{a},\psi_{b}\}_{\scriptscriptstyle{PB}}=-i\delta_{ab}\
,\ \{x^{\mu},\psi_{a}\}_{\scriptscriptstyle{PB}}=0.
\end{equation}
Here, following Ref.\cite{bere}, the Poisson brackets of two
dynamical variables $A(x,p,\psi)$ and $B(x,p,\psi)$ are defined by
\begin{equation}
\{A,B\}_{\scriptscriptstyle{PB}}\equiv\frac{\partial A}{\partial
x^{\mu}} \frac{\partial B}{\partial p_{\mu}}-\frac{\partial
A}{\partial p_{\mu}} \frac{\partial B}{\partial x^{\mu}}
-iA\frac{\overleftarrow{\partial}}{\partial
\psi_{a}}\frac{\overrightarrow{\partial}}{\partial \psi_{a}}B.
\end{equation}
Now one can easily verify that the above supersymmetry
transformation, which can equivalently be written as
\begin{equation}\label{ph-N=1-transf}
\delta x^{\mu}=i\frac{\epsilon}{\sqrt{m}} e^{\mu}_{a}(x)\psi_{a}\
,\ \ \delta\psi_{a}=-\frac{\epsilon}{\sqrt{m}}
e^{\mu}_{a}(x)(p_{\mu}-A_{\mu}(x)),
\end{equation}
is generated by the supercharge $Q_{1}\equiv
\frac{i}{\sqrt{m}}\psi_{a}e^{\mu}_{a}(x)(p_{\mu}-A_{\mu}(x))$,
i.e.,
\begin{equation}
\delta x^{\mu}=\{x^{\mu},\epsilon
Q_{1}\}_{\scriptscriptstyle{PB}}\ ,\ \delta
\psi_{a}=\{\psi_{a},\epsilon Q_{1}\}_{\scriptscriptstyle{PB}}.
\end{equation}
Note that $\{H,Q_{1}\}_{\scriptscriptstyle{PB}}=0$, and hence the
charge $Q_{1}$ is conserved. Also the Hamiltonian
(\ref{curved-ham}) can be expressed as
\begin{equation}\label{susy-ham}
H=-\frac{i}{2}\{Q_{1},Q_{1}\}_{\scriptscriptstyle{PB}}.
\end{equation}

Now, to exhibit the $N=2$ supersymmetry for our system, define
\begin{equation}
\psi\equiv\frac{1}{\sqrt{2}}(\psi_{1}\!-\!i\psi_{2})\ ,\
e^{\mu}(x)\equiv\frac{1}{\sqrt{2}}(e^{\mu}_{1}(x)\!-\!ie^{\mu}_{2}(x)),
\end{equation}
where the subscript $1,2$ refer to components relative to the
local orthonormal frame. Also let us denote the complex conjugates
of $\psi$, $e^{\mu}(x)$ by $\bar{\psi}$, $\bar{e}^{\mu}(x)$. Then
it is shown by direct calculations that a pair of complex
supercharges
\begin{equation}\label{N=2-charge}
Q\equiv
\frac{i}{\sqrt{m}}\psi\bar{e}^{\mu}(x)(p_{\mu}-A_{\mu}(x))\ ,\
\bar{Q}\equiv-\frac{i}{\sqrt{m}}\bar{\psi}e^{\mu}(x)(p_{\mu}-A_{\mu}(x))
\end{equation}
satisfy the following $N=2$ supersymmetry algebra:
\begin{subeqnarray}
&\{Q,Q\}_{\scriptscriptstyle{PB}}=\{\bar{Q},\bar{Q}\}_{\scriptscriptstyle{PB}}=0&\slabel{null}\\
&i\{Q,\bar{Q}\}_{\scriptscriptstyle{PB}}=H,&\slabel{N=2-ham}\\
&\{H,Q\}_{\scriptscriptstyle{PB}}=\{H,\bar{Q}\}_{\scriptscriptstyle{PB}}=0.&\slabel{conser}
\end{subeqnarray}
Note that $Q_{1}=(Q-\bar{Q})$, and (\ref{N=2-ham}) is just a
rewriting of (\ref{susy-ham}).

For a general dynamical variable $A$, the $N=2$ supersymmetry
transformation can be represented by $\delta A=\{A,\varepsilon Q +
\bar{Q}\bar{\varepsilon}\}_{\scriptscriptstyle{PB}}$, where
$\varepsilon$ is a \textit{complex} Grassmannian parameter.
Especially, for $x^{\mu}$ and $\psi$, the transformation reads
\begin{equation}\label{general-transf}
\delta
x^{\mu}=\frac{i}{\sqrt{m}}(\varepsilon\psi\bar{e}^{\mu}(x)+\bar{\varepsilon}\bar{\psi}e^{\mu}(x))\
,\
\delta\psi=-\frac{1}{\sqrt{m}}\bar{\varepsilon}e^{\mu}(p_{\mu}-A_{\mu}(x)).
\end{equation}
This reduces to (\ref{ph-N=1-transf}) if $\varepsilon$ is taken to
be real Grassmannian (i.e., $\varepsilon=\epsilon$). With
$\varepsilon=i\epsilon$, on the other hand, we obtain from
(\ref{general-transf}) the second supersymmetry transformation
\begin{equation}
\delta
x^{\mu}=\frac{i\epsilon}{\sqrt{m}}\epsilon_{ab}e^{\mu}_{a}(x)\psi_{b}\
,\
\delta\psi_{a}=\frac{\epsilon}{\sqrt{m}}\epsilon_{ab}e^{\mu}_{b}(x)(p_{\mu}-A_{\mu}(x)).
\end{equation}
We remark that, for the Lagrangian (\ref{curved-lag}), this second
supersymmetry can be described by the transformation
(cf.(\ref{x-transf-1}) and (\ref{psi-transf-1'}))
\begin{equation}
\delta
x^{\mu}=\frac{i\epsilon}{\sqrt{m}}\epsilon_{ab}e^{\mu}_{a}(x)\psi_{b}\
,\ \delta\psi_{a}=\frac{\epsilon}{\sqrt{m}}\epsilon_{ab}
 e_{\mu b}(x)(m\dot{x}^{\mu}-\omega^{\mu}(x)S(\psi)).
\end{equation}
For $\psi^{\mu}=e^{\mu}_{a}(x)\psi_{a}$, this leads to the
transformation
$\delta\psi^{\mu}=\frac{\epsilon}{\sqrt{m}}(\epsilon^{\mu}_{\
\nu}(x)m\dot{x}^{\nu}-\Gamma^{\mu}_{\
\nu\rho}(x)g^{\nu\rho}(x)S(\psi))$, where
$\epsilon^{\mu}_{\nu}(x)\equiv\epsilon_{ab}e^{\mu}_{a}(x)e_{\nu b
}(x)$ and $\Gamma^{\mu}_{\ \nu\rho}(x)$ is the Christoffel symbol.
[Note that $\Gamma^{\mu}_{\ \nu\rho}=\frac{1}{2}
g^{\mu\lambda}(\partial_{\nu}g_{\rho\lambda}
+\partial_{\rho}g_{\nu\lambda}-\partial_{\lambda}g_{\nu\rho})$ and
$\omega_{\mu ab}(x)e^{\nu}_{b}(x)=-\partial_{\mu}e^{\nu}_{a}(x)
 -\Gamma^{\nu}_{\ \mu\rho}(x)e^{\rho}_{a}(x)$.]

\section{Quantum Theory with N=2 Supersymmetry}

When the classical Hamiltonian is given as in (\ref{curved-ham}),
the purpose of this section is to set up the corresponding quantum
theory, with the $N=2$ supersymmetry realized by appropriate
Hilbert-space operators. In the Schr\"{o}dinger picture, basic
dynamical variables, i.e., $x^{\mu}$, $p_{\nu}$ and $\psi_{a}$
become Hermitian operators satisfying the algebraic relations
\begin{equation}
\begin{array}{l}
[x^{\mu},p_{\nu}]=i\delta^{\mu}_{\nu}\ ,\
\{\psi_{a},\psi_{b}\}=\delta_{ab},\hspace{0.2cm}\\

[x^{\mu},x^{\nu}]=[p_{\mu},p_{\nu}]=[x^{\mu},\psi_{a}]=[p_{\mu},\psi_{a}]=0,
\end{array}
\end{equation}
where the brackets [ ] and \{ \} denote commutator and
anticommutator, respectively. Note that if the non-Hermitian
operator $\psi=\frac{1}{\sqrt{2}}(\psi_{1}\!-\!i\psi_{2})$ and its
conjugate $\bar{\psi}$ are used, the relations
$\{\psi_{a},\psi_{b}\}=\delta_{ab}$ are equivalent to the
followings: $\{\psi,\bar{\psi}\}=1$ and
$\psi^{2}=\bar{\psi}^{2}=0$. In order to secure the desired
supersymmetric system, we must here require that the quantum
Hamiltonian operator be expressed as
\begin{equation}
H=\{Q,\bar{Q}\}
\end{equation}
with a suitably chosen supercharge operator $Q$ (which is of
course related to the corresponding classical expression in
(\ref{N=2-charge})). Hence, in our discussions below, we are going
to pay much attention to the construction of the supercharge
operators.

To find the quantum operators corresponding to the classical
expressions (\ref{N=2-charge}), one will have to settle first the
operator ordering problem concerning $p_{\mu}$, $e^{\mu}(x)$ and
$\bar{e}^{\mu}(x)$. Following Ref.[21], we will adopt the
Weyl-ordered form, which in the present case yields the operator
\begin{equation}\label{flat-charge-1}
\begin{array}{c}
Q=\frac{i}{\sqrt{m}}\psi\left[\textstyle{\frac{1}{2}}\{\bar{e}^{\mu}(x),p_{\mu}\}\!
-\!\bar{e}^{\mu}(x)A_{\mu}(x)\right]\hspace{2cm}\\
*\hspace{2.5cm}=\frac{i}{\sqrt{m}}\psi\bar{e}^{\mu}(x)\!\left[
 \ p_{\mu}\!-\!A_{\mu}(x)\!+\!\textstyle{\frac{1}{2}}\omega_{\mu}(x)
\!+\!\textstyle{\frac{i}{2}}\Gamma^{\nu}_{\ \nu\mu}(x)\right].
\end{array}
\end{equation}
Here, to obtain the second form, we have assumed that $p_{\mu}$
has the usual differential operator realization, i.e.,
$p_{\mu}=-i\partial_{\mu}$, and used the relation
$\partial_{\mu}\bar{e}^{\nu}(x)=i\omega_{\mu}(x)\bar{e}^{\nu}(x)-\Gamma^{\nu}_{\
\mu\rho}(x)\bar{e}^{\rho}(x)$, which follows from the
corresponding relation for $\partial_{\mu}e^{\nu}_{a}(x)$. The
operator $\bar{Q}$ is then given by Hermitian conjugation, and one
can express the result as
\begin{equation}\label{flat-charge-2}
\bar{Q}=-\frac{i}{\sqrt{m}}\bar{\psi}e^{\mu}(x)\!\left[
 \ p_{\mu}\!-\!A_{\mu}(x)\!-\!\textstyle{\frac{1}{2}}\omega_{\mu}(x)
\!+\!\textstyle{\frac{i}{2}}\Gamma^{\nu}_{\ \nu\mu}(x)\right].
\end{equation}
The way that the Christoffel symbols and spin connection enter
these expressions may appear somewhat perplexing. But it has a
simple geometrical origin, which can be seen if one considers the
Hilbert space structure of wave functions defined on a curved
manifold. This aspect is discussed below.

The differential operators $p_{\mu}=-i\partial_{\mu}$ provide a
faithful representation of the \textit{Hermitian} momentum
operators if the inner product of two state vectors
$|\Psi_{1}\rangle$ and $|\Psi_{2}\rangle$, with corresponding wave
functions $\Psi_{1}(x)$ and $\Psi_{2}(x)$, is taken as
\begin{equation}\label{flat-norm}
\langle\Psi_{1}|\Psi_{2}\rangle=\int
d^{2}x\Psi_{1}^{\ast}(x)\Psi_{2}(x).
\end{equation}
Quantum operators obtained by Weyl-ordering the classical
expressions may well act on the states  the norm of which are
defined with the help of this inner product. In a curved surface,
however, we have more convenient inner product in the one using
the invariant volume measure $\sqrt{g}\ d^{2}x$, i.e.,
\begin{equation}\label{curved-norm}
\langle\Psi_{1}|\Psi_{2}\rangle=\int d^{2}x\sqrt{g(x)}
\hat{\Psi}_{1}^{\ast}(x)\hat{\Psi}_{2}(x),
\end{equation}
where $\hat{\Psi}(x)$ denotes the `invariant' wave function
associated with the ket $|\Psi\rangle$. Evidently, these
differently normalized wavefunctions are related by
\begin{equation}\label{waveftn-transf}
\hat{\Psi}(x)=g^{\!-\!1\!/\!4\!}(x)\Psi(x).
\end{equation}
Then, given a differential operator realization $\Omega$ in the
Hilbert space equipped with the inner product (\ref{flat-norm}),
we are led to consider, if the inner product (\ref{curved-norm})
is assumed instead, the differential operator realization
\begin{equation}
\hat{\Omega}=g^{\!-\!1\!/\!4\!}(x)\Omega g^{\!1\!/\!4\!}(x).
\end{equation}
This becomes nontrivial only for that containing the momentum
operator, with $p_{\mu}=-i\partial_{\mu}$ and
\begin{equation}\label{mom-transf}
\begin{array}{l}
\hat{p}_{\mu}=g^{\!-\!1\!/\!4\!}(x)p_{\mu}g^{\!1\!/\!4\!}(x)
=p_{\mu}-\frac{i}{4}\partial_{\mu}[\log
g(x)]\vspace{0.2cm}\\
*\hspace{0.5cm}=p_{\mu}-\frac{i}{2}\Gamma^{\nu}_{\
\nu\mu}(x).\hspace{3cm}
\end{array}
\end{equation}
For the wave function $\hat{\Psi}(x)$, the differential operators
$\hat{p}_{\mu}$ should be used to represent the Hermitian momentum
operators. Now, noting (thanks to the anticommutation relations
for $\psi$, $\bar{\psi}$)) that $\pm\frac{1}{2}\omega_{\mu}(x)$
may be replaced by
$\frac{1}{2}\omega_{\mu}(x)[\bar{\psi},\psi]=\omega_{\mu}(x)S(\psi)$,
we may write the supercharge operators $\hat{Q}\
(=g^{\!-\!1\!/\!4\!}(x)Qg^{\!1\!/\!4\!}(x))$ and $\hat{\bar{Q}}\
(=g^{\!-\!1\!/\!4\!}(x)\bar{Q}g^{\!1\!/\!4\!}(x))$ as
\begin{subeqnarray}
&\hat{Q}=\frac{i}{\sqrt{m}}\psi\bar{e}^{\mu}(x)\!\left[
 \ p_{\mu}\!-\!A_{\mu}(x)\!+
 \!\omega_{\mu}(x)S(\psi)\right],&\label{curved-charge-1'}\\
 &\hat{\bar{Q}}=-\frac{i}{\sqrt{m}}\bar{\psi}e^{\mu}(x)\!\left[
 \ p_{\mu}\!-\!A_{\mu}(x)\!+\!\omega_{\mu}(x)S(\psi)
\right].&\label{curved-charge-2'}
\end{subeqnarray}
These forms now look quite reasonable. But, in these quantum
expressions(cf. the classical ones in (\ref{N=2-charge})), notice
the presence of the term coupling the spin connection to the spin
operator $S(\psi)$.

From now on we shall assume the inner product given by
(\ref{curved-norm}) for our wave functions and use the expressions
in (3.10a,b) for the supercharge operators (with the hats over
$Q$, $\bar{Q}$ now suppressed, but retaining the notation
$p_{\mu}=-i\partial_{\mu}$). With the first supercharge operator
$Q_{1}=\frac{1}{i}(Q-\bar{Q})$ having the explicit form
\begin{equation}
Q_{1}=\frac{1}{\sqrt{m}}\psi_{a}e^{\mu}_{a}(x)[p_{\mu}-A_{\mu}(x)+\omega_{\mu}(x)S(\psi)],
\end{equation}
the Hamiltonian operator can then be identified with
$H=Q_{1}^{2}$. To write this operator in more familiar form, let
us employ the $2\times2$ matrix representation for the fermionic
operators $\psi_{a}=\frac{1}{\sqrt{2}}\sigma_{a}\ (a=1,2)$, so
that we may have $S(\psi)=\frac{1}{2}\sigma_{3}$. Then the
supercharge $Q_{1}$, acting on a two-component spinor wave
function, acquires the form
\begin{equation}
Q_{1}=\frac{1}{\sqrt{2m}}\sigma_{a}e^{\mu}_{a}(x)[-i\partial_{\mu}
-A_{\mu}+\frac{1}{2}\sigma_{3}\omega_{\mu}(x)]\equiv
\frac{1}{\sqrt{2m}}\sigma_{a}e^{\mu}_{a}(x)(-iD_{\mu}),
\end{equation}
where $D_{\mu}$ is the covariant derivative which include both the
gauge- and spin- connection terms. We now find the Hamiltonian
operator
\begin{equation}\label{sch-ham}
\begin{array}{l}
{\displaystyle{H=-\frac{1}{2m}[\sigma_{a}e^{\mu}_{a}(x)D_{\mu}]^{2}}}\hspace{4.6cm}\vspace{0.15cm}\\
*\hspace{0.5cm}{\displaystyle{=-\frac{1}{2m}\frac{1}{\sqrt{g}}D_{\mu}\sqrt{g}g^{\mu\nu}D_{\nu}
-\frac{1}{2m}\sigma_{3}B(x)+\frac{1}{8m}R(x),}}
\end{array}
\end{equation}
where
$R(x)\equiv\epsilon_{ab}e^{\mu}_{a}(x)e^{\nu}_{a}(x)R_{\mu\nu}(x)$,
with
$R_{\mu\nu}(x)=\partial_{\mu}\omega_{\nu}(x)-\partial_{\nu}\omega_{\mu}(x)$,
is the Ricci scalar for the surface. [Note that, in two
dimensions, the Riemann curvature contains only the pieces linear
in  $\omega_{\mu ab}$]. As one can easily verify, the operators
$H$, $Q$ and $\bar{Q}$ we have just constructed satisfy the $N=2$
supersymmetry algebra:
\begin{subeqnarray}
&&\{Q,Q\}=\{\bar{Q},\bar{Q}\}=0,\\
&&\{Q,\bar{Q}\}=H,\\
&&[H,Q]=[H,\bar{Q}]=0.
\end{subeqnarray}
We also remark that, with $\psi_{a}=\frac{1}{\sqrt{2}}\sigma_{a}$,
matrices for $\psi$ and $\bar{\psi}$ read $\psi=\pmatrix{ 0 & 0
\cr 1 & 0 } \equiv\sigma_{-}$ and $\bar{\psi}=\pmatrix{ 0 & 1 \cr
0 & 0 }\equiv\sigma_{+}$.

\vspace{0.3cm}
 The matrix Hamiltonian
(\ref{sch-ham}), the curved-space generalization of the flat-space
Pauli Hamiltonian in (\ref{flat-ham}), should be relevant
apparently for a spin-$1/2$ particle. But, in nonrelativistic
quantum mechanics, `spin' is nothing more than another
\textit{internal} degrees of freedom; in this light, the above
Hamiltonian may well be reinterpreted just as that containing a
pair of Schr\"{o}dinger Hamiltonains considered for a scalar
particle. For the purpose, we employ the above $2\times2$ matrix
representations for $\psi$, $\bar{\psi}$ with the supercharge
operators in (3.10a,b), to write
\begin{equation}\label{sc-fac}
\begin{array}{l}
H=Q\bar{Q}+\bar{Q}Q\hspace{5cm}\\
*\hspace{0.5cm}=\frac{1}{m
}\left(\begin{array}{cc}\pi_{-}\pi_{+}&0\\0&\pi_{+}\pi_{-}\end{array}\right)\equiv
\left(\begin{array}{cc}H_{+}&0\\0&H_{-}\end{array}\right),
\end{array}
\end{equation}
where we have defined
\begin{equation}\label{ladder}
\pi_{+}\equiv
\bar{e}^{\mu}(x)[p_{\mu}\!-\!A_{\mu}(x)\!+\!\frac{1}{2}\omega_{\mu}(x)]\
,\ \pi_{-}\equiv
e^{\mu}(x)[p_{\mu}\!-\!A_{\mu}(x)\!-\!\frac{1}{2}\omega_{\mu}(x)].
\end{equation}

In (\ref{sc-fac}) we have two Hamiltonians
$H_{\pm}=\frac{1}{m}\pi_{\mp}\pi_{\pm}$ connected by supersymmetry
transformations. As we set
${\mathcal{A}}^{+}_{\mu}=A_{\mu}-\frac{1}{2}\omega_{\mu}$, it is
also evident from (\ref{sch-ham}) that the Hamiltonian $H_{+}$ can
be expressed by the form
\begin{equation}\label{sc-ham-1}
H_{+}=
-\frac{1}{2m}\frac{1}{\sqrt{g}}D^{+}_{\mu}\sqrt{g}g^{\mu\nu}D^{+}_{\nu}-\frac{1}{2m}{\mathcal{B}}^{+}(x),
\end{equation}
where $D^{+}_{\mu}=\partial_{\mu}-i{\mathcal{A}}^{+}_{\mu}(x)$,
and ${\mathcal{B}^{+}}$ is the magnetic field obtained from the
vector potential ${\mathcal{A}}^{+}_{\mu}$. Similarly, writing
${\mathcal{A}}^{-}_{\mu}\equiv A_{\mu}+\frac{1}{2}\omega_{\mu}$,
we can express the second Hamiltonian $H_{-}$ as
\begin{equation}\label{sc-ham-2}
H_{-}=
-\frac{1}{2m}\frac{1}{\sqrt{g}}D^{-}_{\mu}\sqrt{g}g^{\mu\nu}D^{-}_{\nu}+\frac{1}{2m}{\mathcal{B}}^{-}(x),
\end{equation}
with $D^{-}_{\mu}=\partial_{\mu}-i{\mathcal{A}}^{-}_{\mu}(x)$.
From this discussion it follows that our system contains a
supersymmetric pair of curved-space scalar Hamiltonians, i.e.,
$H_{+}$ and $H_{-}$, with $H_{+}(H_{-})$ defined in the
simultaneous presence of the vector potential
${\mathcal{A}}^{+}_{\mu}({\mathcal{A}}^{-}_{\mu})$ and the scalar
potential
$-\frac{1}{2m}{\mathcal{B}^{+}}({+\frac{1}{2m}\mathcal{B}^{-}})$.
The problem based on this view and that based on the spin-$1/2$
particle interpretation are \textit{mathematically} equivalent;
one can go from one problem to the other by identifying things
like potentials differently. It also follows, from the
supersymmetry algebra, that the above Hamiltonian $H_{\pm}$ admit
only non-negative eigenvalues.

\section{Complete Energy Eigenstates for Shape-Invariant Systems}

The supersymmetric pair Hamiltonian structure found in the
previous section can be used to find the full Hilbert space (i.e.,
complete energy eigenstates) for the Landau system defined on the
two-sphere or on the hyperbolic plane. This becomes possible
because, in the case of a \textit{constant} magnetic field present
over a \textit{constant-curvature} surface, the corresponding
Hamiltonian acquires additionally the so-called shape-invariance
porperty [11,12] which allows one to develop simple algebraic
methods to find the spectrum and energy eigenstates. [But, in
contrast to the one-dimensional examples considered in Refs.
[11,12], we here have two-dimensional shape-invariant systems]. In
what follows, we will assume that the Hamiltonian we wish to
really study is $H_{+}$ (as given by (\ref{sc-ham-1})), with
${\mathcal{A}}^{+}_{\mu}(x)$ now renamed as $A_{\mu}(x)$ (and,
correspondingly, ${\mathcal{A}}^{-}_{\mu}(x)$ as
$A_{\mu}(x)+\omega_{\mu}(x)$). Also, in the present situation
where $B(x)$ and $R(x)$ are constant, $F_{\mu\nu}(x)\
(=\partial_{\mu}A_{\nu}(x)-\partial_{\nu}A_{\mu}(x))$ and
$R_{\mu\nu}(x)\
(=\partial_{\mu}\omega_{\nu}(x)-\partial_{\nu}\omega_{\mu}(x))$
are both proportional to $\epsilon_{\mu\nu}\
(=e^{a}_{\mu}e^{b}_{\nu}\epsilon_{ab})$ and so we may write, with
a judicious choice of the zweibeins,
\begin{equation}\label{const-vec-pot}
A_{\mu}(x)=\gamma\omega_{\mu}(x),
\end{equation}
where $\gamma$ is a constant. We remark that $\gamma$ can be any
real number in the case of the hyperbolic plane. But, in the
two-sphere case, the famous Dirac quantization condition[22] for a
globally well-defined one-form potential demands that $\gamma$ be
restricted to half-integral values.

Let us denote the vector potential in (\ref{const-vec-pot}) by
$A^{\gamma}_{\mu}(x)$, and the corresponding magnetic field
strength by $B^{\gamma}$ ($=\frac{1}{2}\gamma R$, if $R$ is the
constant scalar curvature of the surface). We write the
Hamiltonian $H_{+}$ (in (\ref{sc-ham-1})) with
$A^{\gamma}_{\mu}(x)$ taking the place of
${\mathcal{A}}^{+}_{\mu}(x)$ as ${\mathcal{H}}_{\gamma}$, i.e.,
\begin{equation}
{\mathcal{H}}_{\gamma}=-\frac{1}{2m}\frac{1}{\sqrt{g(x)}}
D^{\gamma}_{\mu}\sqrt{g(x)}g^{\mu\nu}(x)
D^{\gamma}_{\nu}-\frac{1}{2m}B^{\gamma},\ \
(D^{\gamma}_{\mu}\equiv\partial_{\mu}-iA^{\gamma}_{\mu}).
\end{equation}
But for the trivial additive constant $\frac{1}{2m}B^{\gamma}$,
${\mathcal{H}}_{\gamma}$ is really the Landau Hamiltonian  on the
surface. We know that ${\mathcal{H}}_{\gamma}$ may also be written
in the form
\begin{equation}\label{fac-1}
{\mathcal{H}}_{\gamma}=\frac{1}{m}\pi_{\gamma+1}\bar{\pi}_{\gamma},
\end{equation}
where we defined the operators $\bar{\pi}_{\gamma},\
\pi_{\gamma+1}$ according to (see (\ref{ladder}))
\begin{equation}
\pi_{\gamma}\equiv e^{\mu}(x)[p_{\mu}-A^{\gamma}_{\mu}(x)]\ ,\ \
\bar{\pi}_{\gamma}\equiv\bar{e}^{\mu}(x)[p_{\mu}-A^{\gamma}_{\mu}(x)]
\end{equation}
and used the fact that
$A^{\gamma+1}_{\mu}(x)=A^{\gamma}_{\mu}(x)+\omega_{\mu}(x)$. On
the other hand, from (\ref{sc-ham-2}) and the relation
$H_{-}=\frac{1}{m}\pi_{+}\pi_{-}$, the superpartner Hamiltonian
related to ${\mathcal{H}}_{\gamma}$ equals
\begin{equation}\label{fac-2}
\begin{array}{l}
\displaystyle{\frac{1}{m}\bar{\pi}_{\gamma}\pi_{\gamma+1}=-\frac{1}{2m}\frac{1}{\sqrt{g(x)}}D^{\gamma+1}_{\mu}
\sqrt{g(x)}g^{\mu\nu}(x)D^{\gamma+1}_{\nu}+\frac{1}{2m}B^{\gamma+1}}\\
*\hspace{1.7cm}={\mathcal{H}}_{\gamma+1}+\frac{1}{m}B^{\gamma+1},\hspace{3.7cm}
\end{array}
\end{equation}
(or
$\frac{1}{m}\bar{\pi}_{\gamma-1}\pi_{\gamma}={\mathcal{H}}_{\gamma}+\frac{1}{m}B^{\gamma}$).
Now suppose that $|E\rangle^{\gamma}$ is an eigenstate of
${\mathcal{H}}_{\gamma}$ with eigenvalue $E$. Then, from
(\ref{fac-1}) and (\ref{fac-2}), we are led to the relation
\begin{equation}\label{shape-inv-rel-1}
\begin{array}{l}
{\mathcal{H}}_{\gamma+1}\bar{\pi}_{\gamma}|E\rangle^{\gamma}=
(\frac{1}{m}\bar{\pi}_{\gamma}\pi_{\gamma+1}-\frac{1}{m}B^{\gamma+1})\bar{\pi}_{\gamma}|E\rangle^{\gamma}=
\bar{\pi}_{\gamma}({\mathcal{H}}_{\gamma}-\frac{1}{m}B^{\gamma+1})|E\rangle^{\gamma}
\\
*\hspace{2.2cm}=(E-\frac{1}{m}B^{\gamma+1})\bar{\pi}_{\gamma}|E\rangle^{\gamma},
\end{array}
\end{equation}
i.e., $\bar{\pi}_{\gamma}|E\rangle^{\gamma}$ is an eigenstate of
${\mathcal{H}}_{\gamma+1}$ with energy
$E-\frac{1}{m}B^{\gamma+1}$. Analogously, for
$\pi_{\gamma}|E\rangle^{\gamma}$, we find
\begin{equation}\label{shape-inv-rel-2}
{\mathcal{H}}_{\gamma-1}\pi_{\gamma}|E\rangle^{\gamma}=
\frac{1}{m}\pi_{\gamma}\bar{\pi}_{\gamma-1}\pi_{\gamma}|E\rangle^{\gamma}
=(E+\frac{1}{m}B^{\gamma})\pi_{\gamma}|E\rangle^{\gamma},
\end{equation}
i.e., $\pi_{\gamma}|E\rangle^{\gamma}$ is an eigenstate of
${\mathcal{H}}_{\gamma-1}$ with eigenvalue
$E+\frac{1}{m}B^{\gamma}$.

The particular structure shown in (\ref{fac-1}) and (\ref{fac-2})
describes the shape-invariance property of our Hamiltonian
${\mathcal{H}}_{\gamma}$ (with parameter $\gamma$), which we will
utilize to find the complete energy eigenstates. Without loss of
generality we may here assume $B^{\gamma}=\frac{1}{2}\gamma R>0$;
this corresponds to the choice $\gamma>0$ if $R>0$ (i.e., in the
case of the two-sphere), and $\gamma<0$ if $R<0$ (the hyperbolic
plane). Now, for any given eigenstate $|E\rangle^{\gamma}$ of
${\mathcal{H}}_{\gamma}$, we know from this shape-invariance
property that
$\bar{\pi}_{\gamma+n-1}\bar{\pi}_{\gamma+n-2}\cdots\bar{\pi}_{\gamma}|E\rangle^{\gamma}$,
if it is not a zero vector, should correspond to an eigenstate of
${\mathcal{H}}_{\gamma+n}$ with energy equal to
$E-\frac{1}{m}B^{\gamma+1}\cdots-\frac{1}{m}B^{\gamma+n}$. In the
case of the two-sphere where
$B^{\gamma+n}=\frac{1}{2}(\gamma+n)R>0$ for all $n=0,1,2,\cdots$,
this procedure of generating new states must stop after applying a
finite number of $\bar{\pi}$'s (since the eigenvalue of
${\mathcal{H}}_{\gamma+n}$ for any $n\geq 0$ cannot be negative);
i.e., for any $|E\rangle^{\gamma}$, there exists some non-negative
integer $s$ such that
\begin{equation}\label{annih-rel}
\bar{\pi}_{\gamma+s}\bar{\pi}_{\gamma+s-1}\cdots\bar{\pi}_{\gamma}|E\rangle^{\gamma}=0,\hspace{1cm}
\end{equation}
while
$\bar{\pi}_{\gamma+s-1}\bar{\pi}_{\gamma+s-2}\cdots\bar{\pi}_{\gamma}|E\rangle^{\gamma}\neq
0$. When (\ref{annih-rel}) holds,
$\bar{\pi}_{\gamma+s-1}\bar{\pi}_{\gamma+s-2}\cdots\bar{\pi}_{\gamma}|E\rangle^{\gamma}$
corresponds to a \textit{zero-energy} eigenstate of
${\mathcal{H}}_{\gamma+s}=\pi_{\gamma+s+1}\bar{\pi}_{\gamma+s}$
(i.e.,$\bar{\pi}_{\gamma+s-1}$
$\bar{\pi}_{\gamma+s-2}\cdots\bar{\pi}_{\gamma}
|E\rangle^{\gamma}\propto|0\rangle^{\gamma+s}$) and therefore
\begin{equation}\label{discr-energy}
E-\frac{1}{m}B^{\gamma+1}-\cdots-\frac{1}{m}B^{\gamma+s}=0.
\end{equation}
The relation (\ref{discr-energy}) determines the eigenvalue $E$ of
${\mathcal{H}}_{\gamma}$ in terms of the \textit{quantum number}
$s$, showing also that we have a discrete spectrum only. In the
case of the hyperbolic plane (with $R<0$ and $\gamma<0$), however,
the situation is somewhat different since
$B^{\gamma+n}=\frac{1}{2}\gamma R +\frac{1}{2}nR$ changes sign
from the positive to the negative for large enough $n$. In the
latter case there exist not only discrete energy levels
(determined by the condition like (\ref{discr-energy})) but also a
continuous spectrum for energy exceeding some critical value[13].
We shall below construct the complete energy eigenstates for the
two cases separately.\vspace{0.5cm}

{\bf A.\textit{Complete Energy Eigenstates on the
Two-sphere}\vspace{0.35cm}}

Here, as we remarked already, $\gamma(>0)$ must be half integral.
Since there exists a discrete spectrum only, we may denote the
eigenstates of ${\mathcal{H}}_{\gamma}$ by
$|E_{s}\rangle^{\gamma}$. Allowed values of $E_{s}$ --- the
eigenvalue spectrum of ${\mathcal{H}}_{\gamma}$ --- are determined
by solving (\ref{discr-energy}):
\begin{equation}\label{sphere-spectrum}
E_{s}=\frac{1}{m}\sum_{k=1}^{s} B^{\gamma+k}
=\frac{R}{4m}s(2\gamma+s+1),\ \ \ (s=0,1,2,\cdots).
\end{equation}
To find the full Hilbert space, we may begin with determining the
general wave functions corresponding to the states
$|0\rangle^{\gamma+s}$. This is a relatively easy task, for they
have the property of being annihilated by the first-order
operator, that is,
\begin{equation}\label{zero-state}
\bar{\pi}_{\gamma+s}|0\rangle^{\gamma+s}=0.
\end{equation}
Once they are known, we may use (\ref{shape-inv-rel-2}): this
relation tells us that $|E_{s}\rangle^{\gamma}$ may be obtained
simply by acting
$\pi_{\gamma+1}\pi_{\gamma+2}\cdots\pi_{\gamma+s}$ on
$|0\rangle^{\gamma+s}$. Indeed, assuming that
$|0\rangle^{\gamma+s}$ is normalized, we can obtain the correctly
normalized states $|E_{s}\rangle^{\gamma}$ through\vspace{0.2cm}
\begin{equation}\label{normalise-1}
|E_{s}\rangle^{\gamma}=\sqrt{\frac{\Gamma(2\gamma+s+1)}{\Gamma(s+1)\Gamma(2\gamma+2s+1)}}
\pi_{\gamma+1}\pi_{\gamma+2}\cdots\pi_{\gamma+s}|0\rangle^{\gamma+s}.
\end{equation}\vspace{0.2cm}
(One may recall our earlier finding $|0\rangle^{\gamma+s}\propto
\bar{\pi}_{\gamma+s-1}\bar{\pi}_{\gamma+s-2}$
$\cdots\bar{\pi}_{\gamma}|E_{s}\rangle^{\gamma}$). To fix the
normalization constant in (\ref{normalise-1}), we have made use of
the relations in (\ref{fac-1})-(\ref{shape-inv-rel-2}).

Because of the shape-invariance in the system, one can thus reduce
the problem of finding complete energy eigenfunctions for a given
magnetic field strength essentially to that of finding zero-energy
eigenfunctions for different magnetic field  strengths or to the
analysis of (\ref{zero-state}). For explicit eigenfunctions,
suitable coordinates must be chosen[Remember that our wave
function defines a scalar field under coordinate transformation].
To obtain explicit zero-energy wave functions using spherical
coordinates $\theta,\ \phi$ for instance, we may consider the
one-forms $\textbf{e}^{a}$ given by
\begin{eqnarray}
&&\textbf{e}^{a}=\Lambda(\Theta)^{a}_{\ b}\textbf{e}^{b}_{(0)},\label{rot-zwei}\\
&&\Lambda(\Theta)=\left(\begin{array}{cc}\cos\Theta(\theta,\phi)\ ,&-\sin\Theta(\theta,\phi)\\
\sin\Theta(\theta,\phi)\
,&\cos\Theta(\theta,\phi)\label{rot-matrix}
\end{array}\right)\in SO(2)
\end{eqnarray}
where $\textbf{e}^{a}_{(0)}$ are the usually chosen one-forms
related to the line element
$g_{\mu\nu}dx^{\mu}dx^{\nu}=r^{2}d\theta^{2}+r^2\sin^{2}\theta
d\phi^{2}$ (the constant $r$ here represents the radius of the
sphere), i.e.,
\begin{equation}
\begin{array}{l}
\textbf{e}^{1}_{(0)}\equiv e^{1}_{(0)\theta}d\theta+e^{1}_{(0)\phi}d\phi=rd\theta,\vspace{0.3cm}\\
\textbf{e}^{2}_{(0)}\equiv
e^{2}_{(0)\theta}d\theta+e^{2}_{(0)\phi}d\phi=r\sin\theta d\phi.
\end{array}
\end{equation}
The local frame rotation $\Lambda(\Theta)$ above are to be chosen
such that the vector potentials $A_{\mu}$ as given by
(\ref{const-vec-pot}) may acquire most convenient forms.
Calculating the spin connections using the form (\ref{rot-zwei})
yields
\begin{equation}\label{rot-conn}
\omega_{\mu}=\omega_{(0)\mu}+\partial_{\mu}\Theta
\end{equation}
with $\omega_{(0)\theta}=0$ and $\omega_{(0)\phi}=-\cos\theta$,
and for the scalar curvature we find the value
$R=\frac{2}{r^{2}}$. We now choose $\Theta(\theta,\phi)=\phi$.
Then
\begin{equation}
\omega_{\theta}=0\ ,\ \ \omega_{\phi}=1-\cos\theta,
\end{equation}
and from (\ref{const-vec-pot}) we obtain the familiar
magnetic-monopole vector potentials[22]
\begin{equation}
A_{\theta}=0\ ,\ \ A_{\phi}=\gamma(1-\cos\theta),
\end{equation}
corresponding to the constant magnetic field strength
$B=\frac{\gamma}{r^{2}}$.

Based on the above information, one has the operator
$\bar{\pi}_{\gamma}$ in spherical coordinates expressed by the
form
\begin{equation}\label{sph-ladder-bar}
\bar{\pi}_{\gamma}=-\frac{i}{\sqrt{2}r}e^{i\phi}
\left\{(\partial_{\theta}+\gamma\frac{1-\cos\theta}{\sin\theta})+\frac{i}{\sin\theta}\partial_{\phi}\right\}.
\end{equation}
Evidently, acting on an eigenstate of the angular momentum
$-i\partial_{\phi}$ (with eigenvalue $l=0,\pm1,\pm2,\cdots$), the
operator $\bar{\pi}_{\gamma}$ will have the effect of raising the
value of $l$ by 1. Let
\begin{equation}
\Psi^{\gamma}_{s,l}(\theta,\phi)\equiv\langle\theta,\phi|E_{s},l\rangle^{\gamma}
=\Phi^{\gamma}_{s,l}(\theta)e^{il\phi}
\end{equation}
denote the normalized eigenfunction of ${\mathcal{H}}_{\gamma}$
with energy $E_{s}$ (in (\ref{sphere-spectrum})) and angular
momentum $l$. Then, from (\ref{sph-ladder-bar}),
\begin{equation}\label{sph-bar-ang-decomp}
\bar{\pi}_{\gamma}\Psi^{\gamma}_{s,l}(\theta,\phi)=
[\bar{\pi}^{(l)}_{\gamma}\Psi^{\gamma}_{s,l}(\theta)]e^{i(l+1)\phi},
\end{equation}
where the operator $\bar{\pi}^{(l)}_{\gamma}$ is given as
\begin{equation}\label{sph-ang-bar}
\begin{array}{l}
\bar{\pi}^{(l)}_{\gamma}=-{\displaystyle{\frac{i}{\sqrt{2}r}
(\partial_{\theta}+\gamma\frac{1-\cos\theta}{\sin\theta}-\frac{l}{\sin\theta})}}\
\vspace{0.3cm}\\
\hspace{0.7cm}={\displaystyle{\frac{i}{\sqrt{2}r}(1\!-\!\cos\theta)^{\frac{l+1}{2}}
(1\!+\!\cos\theta)^{\gamma+1-\frac{l+1}{2}}\frac{\vec{\partial}}{\partial(\cos\theta)}}}
(1\!-\!\cos\theta)^{-\frac{l}{2}}(1\!+\!\cos\theta)^{-\gamma+\frac{l}{2}}.
\end{array}
\end{equation}

Now, if
$\Psi^{\gamma+s}_{(0),l}(\theta,\phi)=\Phi^{\gamma+s}_{(0),l}(\theta)e^{il\phi}$
corresponds to a zero-energy state of ${\mathcal{H}_{\gamma+s}}$,
we know from (\ref{zero-state}) and (\ref{sph-bar-ang-decomp})
that $\Phi^{\gamma+s}_{(0),l}(\theta)$ must satisfy the equation
$\bar{\pi}^{(l)}_{\gamma+s}\Phi^{\gamma+s}_{(0),l}(\theta)=0$.
Then, using (\ref{sph-ang-bar}) for $\bar{\pi}^{(l)}_{\gamma+s}$,
it follows that $\Phi^{\gamma+s}_{(0),l}(\theta)$ is proportional
to
$(1\!-\!\cos\theta)^{\frac{l}{2}}(1\!+\!\cos\theta)^{\gamma+s-\frac{l}{2}}$.
From this discussion the desired normalized wave function
$\Psi^{\gamma+s}_{(0),l}(\theta,\phi)$ is found to have the
explicit form
\begin{equation}
\begin{array}{r}
\Psi^{\gamma+s}_{(0),l}(\theta,\phi)=\frac{1}{2^{\scriptscriptstyle{\gamma+s}}}
\sqrt{\frac{\Gamma(2\gamma+2s+2)}{4\pi\Gamma(2\gamma+2s-l+1)\Gamma(l+1)}}
(1\!-\!\cos\theta)^{\frac{l}{2}}(1\!+\!\cos\theta)^{\gamma+s-\frac{l}{2}}e^{il\phi},\\
(l=0,1,\cdots,2\gamma+2s)
\end{array}
\end{equation}
where the restriction on the quantum number $l$ derives from the
normalizability requirement. All that remains is to use
(\ref{normalise-1}) to find the complete normalized eigenfunctions
of ${\mathcal{H}}_{\gamma}$. Here, since the operator
$\pi_{\gamma}$ has the effect of lowering the angular momentum
value by 1, one can facilitate the calculation by introducing the
operator analogous to $\bar{\pi}^{(l)}_{\gamma}$ (in
(\ref{sph-ang-bar})); it equals the expression
\begin{equation}
\pi^{(l)}_{\gamma}={\frac{i}{\sqrt{2}r}(1\!-\!\cos\theta)^{-\frac{l-1}{2}}
(1\!+\!\cos\theta)^{-(\gamma-1)+\frac{l-1}{2}}\frac{\vec{\partial}}{\partial(\cos\theta)}}
(1\!-\!\cos\theta)^{\frac{l}{2}}(1\!+\!\cos\theta)^{\gamma-\frac{l}{2}}.
\end{equation}
The effect of applying
$\pi_{\gamma+1}\cdots\pi_{\gamma+s-1}\pi_{\gamma+s}$ on
$\Psi^{\gamma+s}_{(0),l}(\theta,\phi)=\Phi^{\gamma+s}_{(0),l}(\theta)e^{il\phi}$
(see (\ref{normalise-1})) can then be described by the action of
$\pi^{(l-s)}_{\gamma+1}\cdots\pi^{(l-1)}_{\gamma+s-1}\pi^{(l)}_{\gamma+s}$
on $\Phi^{\gamma+s}_{(0),l}(\theta)$ combined with the change
$e^{il\phi}\rightarrow e^{i(l-s)\phi}$. The result is the
following form for complete energy eigenfunctions(which form
`monopole harmonics'):
\begin{equation}\label{mono-harm}
\begin{array}{r}
\Psi^{\gamma}_{s,l}(\theta,\phi)=\frac{1}{2^{\scriptscriptstyle{\gamma}}}
\sqrt{\frac{(2\gamma+2s+1)\Gamma(s+1)\Gamma(2\gamma+s+1)}{4\pi\Gamma(l+s+1)\Gamma(2\gamma-l+s+1)}}
(1\!-\!\cos\theta)^{\!\frac{l}{2}}(1\!+\!\cos\theta)^{\!\gamma\!-\!\frac{l}{2}}
P^{(l,2\gamma\!-\!l)}_{s}\!(\cos\theta)e^{il\phi},\\
(s=0,1,\cdots; l=-s,-s+1,\cdots,2\gamma+s),\hspace{1cm}
\end{array}
\end{equation}
where
\begin{equation}
P^{(\alpha,\beta)}_{s}(x)=\frac{1}{2^{s}\Gamma(s+1)}(1-x)^{-\alpha}(1+x)^{-\beta}\left(\frac{d}{dx}\right)^{s}
(1-x)^{\alpha+s}(1+x)^{\beta+s}
\end{equation}
is the Jacobi polynomial[23].

If one wishes, the above analysis may be carried out using the
complex coordinates $z,\bar{z}$ which are related to the spherical
coordinates by the stereographic projection
\begin{equation}
z=(\tan\frac{\theta}{2})e^{i\phi}\ \ ,\ \
\bar{z}=(\tan\frac{\theta}{2})e^{-i\phi}.
\end{equation}
Then the operators $\pi_{\gamma},\bar{\pi}_{\gamma}$ become
\begin{eqnarray}
\pi_{\gamma}=-\frac{i}{\sqrt{2}r}\{(1+z\bar{z})\partial_{z}-\gamma\bar{z}\}
=-\frac{i}{\sqrt{2}r}(1+z\bar{z})^{\gamma+1}\vec{\partial}_{z}(1+z\bar{z})^{-\gamma},\\
\bar{\pi}_{\gamma}=-\frac{i}{\sqrt{2}r}\{(1+z\bar{z})\partial_{\bar{z}}+\gamma
z\}
=-\frac{i}{\sqrt{2}r}(1+z\bar{z})^{-\gamma+1}\vec{\partial}_{\bar{z}}(1+z\bar{z})^{\gamma}.
\end{eqnarray}
Since the complete energy eigenfunctions can be found in much the
same way as above (see also Ref.[14]), we shall here give the
final expression only. It reads
\begin{equation}
\Psi^{\gamma}_{s,l}(z,\bar{z})={\textstyle{\sqrt{\frac{(2\gamma+2s+1)\Gamma(2\gamma+s+1)}
{4\pi\Gamma(s+1)\Gamma(l+s+1)\Gamma(2\gamma-l+s+1)}}}}(1+z\bar{z})^{\gamma}
\left[(1+z\bar{z})^{2}\frac{\partial}{\partial
z}\right]^{s}\frac{z^{l+s}}{(1+z\bar{z})^{\scriptscriptstyle{2(\gamma+s)}}},
\end{equation}
or, in terms of the Jacobi function, we have (note that
$\cos\theta=\frac{1-z\bar{z}}{1+z\bar{z}}$)
\begin{equation}
\Psi^{\gamma}_{s,l}(z,\bar{z})={\textstyle{\sqrt{\frac{(2\gamma+2s+1)\Gamma(s+1)\Gamma(2\gamma+s+1)}
{4\pi\Gamma(l+s+1)\Gamma(2\gamma-l+s+1)}}}}
\frac{z^{l}}{(1\!+\!z\bar{z})^{\scriptscriptstyle{\gamma}}}
P^{(l,2\gamma-l)}_{s}\!\left(\frac{1\!-\!z\bar{z}}{1\!+\!z\bar{z}}\right).
\end{equation}
This is nothing but the wavefunction (\ref{mono-harm}) written in
terms of the coordinates $z$ and $\bar{z}$.\vspace{0.25cm}

{\bf B.\textit{Complete Energy Eigenstates on the Hyperbolic
Plane}}\vspace{0.25cm}

When the scalar curvature $R$ is a negative constant, the
eigenstates $|E\rangle^{\gamma}$ of ${\mathcal{H}}_{\gamma}$ (with
$\gamma$ taken to be negative) come in two distinct families. The
first, a discrete family, corresponds to those with energy $E$ not
exceeding the appropriate critical value so that there exists some
non-negative integer $s$ realizing the annihilation condition
(\ref{annih-rel}), and the second to those with energy eigenvalue
exceeding the critical value. One can readily find the critical
value from studying (\ref{discr-energy}). Here, using
$B^{\gamma+k}=-\frac{1}{2}|R|(-|\gamma|+k)$, we have
\begin{equation}
\begin{array}{l}
{\displaystyle
{\frac{1}{m}\sum_{k=1}^{s}}}B^{\gamma+k}=\frac{|R|}{4m}s(2|\gamma|-s-1)\\
*\hspace{2cm}=-\frac{|R|}{4m}[s-(|\gamma|-\frac{1}{2})]^{2}
+\frac{|R|}{4m}(|\gamma|-\frac{1}{2})^{2},
\end{array}
\end{equation}
and hence the right hand side of (\ref{discr-energy}) remains
positive for any $s$-value if
$E>\frac{|R|}{4m}(|\gamma|-\frac{1}{2})^{2}$, i.e., no solution to
(\ref{discr-energy}) exists. On the other hand, for $E$ smaller
than the critical value
$\frac{|R|}{4m}(|\gamma|-\frac{1}{2})^{2}$, the condition
(\ref{discr-energy}) serves to determine the allowed energy
eigenvalues just as in the case of the two-sphere; this yields the
discrete spectrum
\begin{equation}\label{psu-disc-spectrum}
E_{s}=\frac{|R|}{4m}s(2|\gamma|-s-1),\
(s=0,1,2,\cdots,[|\gamma|-\frac{1}{2}]).
\end{equation}
If $E>\frac{|R|}{4m}(|\gamma|-\frac{1}{2})^{2}$, no such
restriction exists and we expect a continuous spectrum[13]. We
shall below find the complete energy eigenfunctions which are
associated with these two distinct families.

For the energy eigenfunctions corresponding to the discrete
spectrum (\ref{psu-disc-spectrum}), we may first find the
zero-energy eigenfunctions of ${\mathcal{H}}_{-|\gamma|+s}$ (i.e.,
those corresponding to the states $|0\rangle^{-|\gamma|+s}$ in the
specific coordinates chosen), and then use the formula analogous
to (\ref{normalise-1}) to obtain the eigenfunctions corresponding
to the states $|E_{s}\rangle^{\gamma}$. Let us choose the
hyperbolic coordinates $\theta,\phi$ (with $0\leq\theta<\infty$,
$0\leq\phi<2\pi$) in terms of which the line element is described
by the Lobachevsky metric
$ds^{2}=r^{2}d\theta^{2}+r^{2}\sinh^{2}\theta d\phi^{2}$ (for a
real constant r)[24]. The one-forms $\textbf{e}^{a}$ may still be
given by (\ref{rot-zwei}), if $\textbf{e}^{a}_{(0)}$ in the
present case are taken to be (cf.(\ref{rot-matrix}))
\begin{equation}
\textbf{e}^{1}_{(0)}=rd\theta\ \ ,\ \
\textbf{e}^{2}_{(0)}=r\sinh\theta d\phi.
\end{equation}
We here have the scalar curvature $R=-\frac{2}{r^{2}}$, and the
vector potentials $A_{\mu}(=-|\gamma|\omega_{\mu})$ assume the
form
\begin{equation}
A_{\theta}=0\ \ ,\ \ A_{\phi}=|\gamma|(\cosh\theta-1).
\end{equation}
The operators $\bar{\pi}_{\gamma},\pi_{\gamma}$ in hyperbolic
coordinates are then found readily. We may also represent the
normalized eigenfunction of ${\mathcal{H}}_{\gamma}$ with energy
$E_{s}$ and angular momentum $l$ by
$\Psi^{\gamma}_{s,l}(\theta,\phi)=\Phi^{\gamma}_{s,l}(\theta)e^{il\phi}$,
and go on to introduce the operators
$\bar{\pi}^{(l)}_{\gamma},\pi^{(l)}_{\gamma}$ as in the previous
two-sphere case:
\begin{eqnarray}
\bar{\pi}^{(l)}_{\gamma}=-\frac{i}{\sqrt{2}r}(\cosh\theta\!-\!1)^{\frac{l+1}{2}}
(\cosh\theta\!+\!1)^{-|\gamma|+1-\frac{l+1}{2}}\frac{\vec{\partial}}{\partial(\cosh\theta)}
(\cosh\theta\!-\!1)^{-\frac{l}{2}}(\cosh\theta\!+\!1)^{|\gamma|+\frac{l}{2}},&&\label{psu-bar-lad}\\
\pi^{(l)}_{\gamma}=-\frac{i}{\sqrt{2}r}(\cosh\theta\!-\!1)^{-\frac{l-1}{2}}
(\cosh\theta\!+\!1)^{|\gamma|+1+\frac{l-1}{2}}\frac{\vec{\partial}}{\partial(\cosh\theta)}
(\cosh\theta\!-\!1)^{\frac{l}{2}}(\cosh\theta\!+\!1)^{-|\gamma|-\frac{l}{2}}.&&\label{psu-lad}
\end{eqnarray}
Then the eigenfunctions $\Psi^{\gamma}_{s,l}(\theta,\phi)$ are
constructed by the methods parallel to those used in the
two-sphere case. The results read
\begin{equation}\label{continuum-eigenfunction}
\begin{array}{r}
\Psi^{\gamma}_{s,l}(\theta,\phi)=\frac{1}{2^{\scriptscriptstyle{-|\gamma|}}}
\sqrt{\frac{(2|\gamma|-2s-1)\Gamma(s+1)\Gamma(2|\gamma|+l-s)}{4\pi\Gamma(l+s+1)\Gamma(2|\gamma|-s)}}
(\cosh\theta\!-\!1)^{\!\frac{l}{2}}(\cosh\theta\!+\!1)^{\!-|\gamma|\!-\!\frac{l}{2}}
\tilde{P}^{(l,-2|\gamma|\!-\!l)}_{s}\!(\cosh\theta)e^{il\phi},\\
(s=0,1,\cdots, [|\gamma|-\frac{1}{2}];
l=-s,-s+1,-s+2,\cdots),\hspace{1cm}
\end{array}
\end{equation}
where $\tilde{P}^{(\alpha,\beta)}_{s}(x)$ is the associated Jacobi
polynomial[23] given by
\begin{equation}
\tilde{P}^{(\alpha,\beta)}_{s}(x)=\frac{1}{2^{s}\Gamma(s+1)}(x-1)^{-\alpha}(x+1)^{-\beta}\left(\frac{d}{dx}\right)^{s}
(x-1)^{\alpha+s}(x+1)^{\beta+s}.
\end{equation}
Note that, in this (noncompact) hyperbolic-plane case, there are
an infinite number of allowed $l$ values for given $s$ and this of
course implies that each discrete energy level comes with infinite
degeneracy here.

To obtain continuum energy eigenfunctions with
$E>\frac{|R|}{4m}(|\gamma|-\frac{1}{2})^{2}$, one can resort to
the analytic continuation procedure with the above expression for
discrete states, following the strategy of Ref.[25] for general
shape-invariant systems. Here it is useful to note that our
eigenfunction in (\ref{continuum-eigenfunction}) can be rewritten
in terms of the hypergeometric function $F(a,b;c;x)$, based on the
known connection[23] between the latter and the (associated)
Jacobi function:
\begin{equation}\label{conti-hypergeo}
\begin{array}{l}
\Psi^{\gamma}_{s,l}(\theta,\phi)\propto
(\cosh\theta-1)^{\frac{l}{2}}(\cosh\theta+1)^{-|\gamma|-\frac{l}{2}}\\
*\hspace{2cm}\times\frac{\Gamma(s+l+1)}{\Gamma(s+1)\Gamma(l+1)}F(-s,-2|\gamma|+s+1;l+1;\frac{1-\cosh\theta}{2})
e^{il\phi}.
\end{array}
\end{equation}
[In this form it should be understood that
$\frac{1}{\Gamma(l+1)}F(-s,-2|\gamma|+s+1;l+1;\frac{1-\cosh\theta}{2})$
for $l=-s,-s+1,\cdots,-1$ are given by the appropriate limiting
expressions as $l$ approaches negative integer values]. The
expression (\ref{conti-hypergeo}) has the following implication:
if one writes
\begin{equation}\label{conti-hypergeo'}
\Psi^{\gamma}_{s,l}(\theta,\phi)=(\cosh\theta-1)^{\frac{l}{2}}(\cosh\theta+1)^{-|\gamma|-\frac{l}{2}}
F^{\gamma}_{s,l}(\theta)e^{il\phi},
\end{equation}
the Schr\"{o}dinger equation for
$\Psi^{\gamma}_{s,l}(\theta,\phi)$ reduces to a hypergeometric
equation for the function $F^{\gamma}_{s,l}(\theta)$. Now, in this
hypergeometric equation for $F^{\gamma}_{s,l}(\theta)$, one might
wish to dispense with $s$ in favor of the energy
$E(s)=\frac{|R|}{4}s(2|\gamma|-s-1)$; the resulting differential
equation would be just the condition on
$F^{\gamma}_{s,l}(\theta,\phi)$ (with $s$ related to E as we
prescribed), for the function $\Psi^{\gamma}_{s,l}(\theta,\phi)$
given by (\ref{conti-hypergeo'}) to describe an energy eigenstate
with energy $E$. This equation should be valid for general energy
eigenvalue $E$, and it is in this context that we can study the
continuum states also within our approach.

For real $E$ but larger than the critical value
$\frac{|R|}{4m}(|\gamma|-\frac{1}{2})^{2}$, we have two complex
roots $s_{\pm}$ to the equation
$\frac{|R|}{4m}s(2|\gamma|-s-1)=E$, viz.,
\begin{equation}\label{conti-complex-para}
s_{{\scriptscriptstyle{\pm}}}(E)=|\gamma|-\frac{1}{2}\pm
i\sqrt{\frac{4mE}{|R|}-(|\gamma|-\frac{1}{2})^{2}}.
\end{equation}
Then, based on the above discussion, we may simply take the
discrete energy eigenfunction $\Psi^{\gamma}_{s,l}(\theta,\phi)$
in (\ref{conti-hypergeo}) but with $s$ replaced by the complex
value $s_{+}(E)$ or $s_{-}(E)$ to obtain the continuum
eigenfunction corresponding to energy $E$. [Here note that the
hypergeometric function can be considered for complex arguments
also]. For given $E$, this does not yield two independent
eigenfunctions, but just one since
$\Psi^{\gamma}_{s_{+}(E),l}(\theta,\phi)=\Psi^{\gamma}_{s_{-}(E),l}(\theta,\phi)$.
Further, one can check explicitly that this function is everywhere
regular for any integer $l$. [The second-order differential
equation for $F^{\gamma}_{s,l}(\theta)$ has another independent
solution for given $E$, but it does not meet the regularity
requirement near $\theta=0$]. Hence we have the desired continuum
eigenfunction with energy
$E>\frac{|R|}{4m}(|\gamma|-\frac{1}{2})^{2}$ represented by
$\Psi^{\gamma}_{s_{\pm}(E),l}(\theta,\phi)\ (l=0, \pm1, \pm2,
\cdots)$, with $s_{\scriptscriptstyle{\pm}}(E)$ specified by
(\ref{conti-complex-para}). This finding agrees with the result
given in Ref.[26] (after making appropriate changes related to
notational differences).

We remark that, in this case also, the energy eigenfunctions may
be expressed using the complex coordinates $z,\ \bar{z}$ which are
related to the above hyperbolic coordinates by
\begin{equation}
z=(\tanh\frac{\theta}{2})e^{i\phi}\ ,\
\bar{z}=(\tanh\frac{\theta}{2})e^{-i\phi}.
\end{equation}
Now, for operators $\pi_{\gamma},\ \bar{\pi}_{\gamma}$, we have
the following forms:
\begin{eqnarray}
\pi_{\gamma}=-\frac{i}{\sqrt{2}r}\{(1-z\bar{z})\partial_{z}-|\gamma|\bar{z}\}
=-\frac{i}{\sqrt{2}r}(1-z\bar{z})^{-|\gamma|+1}\vec{\partial}_{z}(1-z\bar{z})^{|\gamma|},\\
\bar{\pi}_{\gamma}=-\frac{i}{\sqrt{2}r}\{(1-z\bar{z})\partial_{\bar{z}}+|\gamma|z\}
=-\frac{i}{\sqrt{2}r}(1-z\bar{z})^{|\gamma|+1}\vec{\partial}_{\bar{z}}(1-z\bar{z})^{-|\gamma|}.
\end{eqnarray}
Then, going through the steps parallel to the above development,
we have the discrete energy eigenfunctions for instance described
by
\begin{equation}
\Psi^{\gamma}_{s,l}(z,\bar{z})={\textstyle{\sqrt{\frac{(2|\gamma|-2s-1)\Gamma(2|\gamma|+l-s)}
{4\pi\Gamma(s+1)\Gamma(l+s+1)\Gamma(2|\gamma|-s)}}}}(1-z\bar{z})^{-|\gamma|}
\left[(1-z\bar{z})^{2}\frac{\partial}{\partial
z}\right]^{s}z^{l+s}(1-z\bar{z})^{\scriptscriptstyle{2(|\gamma|-s)}}.
\end{equation}
Using the associated Jacobi function, the last expression becomes
(note that $\cosh\theta=\frac{1+z\bar{z}}{1-z\bar{z}}$ here)
\begin{equation}
\Psi^{\gamma}_{s,l}(z,\bar{z})={\textstyle{\sqrt{\frac{(2|\gamma|-2s-1)\Gamma(s+1)\Gamma(2|\gamma|+l-s)}
{4\pi\Gamma(l+s+1)\Gamma(2|\gamma|-s)}}}}z^{l}(1-z\bar{z})^{|\gamma|}
\tilde{P}^{(l,-2|\gamma|-l)}_{s}\!\left(\frac{1\!+\!z\bar{z}}{1\!-\!z\bar{z}}\right),
\end{equation}
and one may consider the analytic continuation of this form to
obtain the continuum energy eigenfunctions as well.

\section{Conclusions}

We have presented a supersymmetry-based approach as a systematic
method for studying dynamics of a charged particle on a curved
surface in the presence of a perpendicular magnetic field. The N=2
supersymmetric system can be given in two equivalent ways, i.e.,
in the form of the curved-space generalization of the usual Pauli
Hamiltonian or by a matrix Hamiltonian involving a pair of scalar
superpartner Hamiltonians defined on the curved surface. In the
cases where the system possesses the shape-invariance property in
addition, complete solutions to the corresponding quantum
mechanical problems can be found with the help of the simple
operator technique analogous to that used to solve the harmonic
oscillator problem. Based on this idea, we have obtained the
energy levels and complete energy eigenfunctions explicitly for
two specially interesting cases --- the Landau Hamiltonian on the
two-sphere and that on the hyperbolic plane. Compared to the case
of the two-sphere where only discrete levels exist, the Landau
Hamiltonian on the hyperbolic plane has complications due to the
existence of additional continuum states. We have obtained the
eigenfunctions appropriate to such continuum states by considering
a suitable analytic continuation with the expression for discrete
states. We hope that our consideration has exposed the power of
the supersymmetry-based approach in studying Landau Hamiltonians
defined on curved surfaces.

\centerline{\bf ACKNOWLEDGMENTS}\vspace{0.5cm}

This work was supported in part by the BK 21 project of the
Ministry of Education, Korea and the Korea Research Foundation
Grant 2001-015-DP0085.

\end{document}